\documentclass[twocolumn,amsmath,amssymb,osajnl]{revtex4}
\usepackage{epsfig}

\begin{document}

\title{Modal cut-off and the $V$--parameter in photonic crystal fibers}

\author{Niels Asger Mortensen and Jacob Riis Folkenberg}
\affiliation{Crystal Fibre A/S, Blokken 84, DK-3460 Birker\o d, Denmark}

\author{Martin D. Nielsen and Kim P. Hansen}
\affiliation{Crystal Fibre A/S, Blokken 84, DK-3460 Birker\o d, Denmark\\COM, Technical University of Denmark, DK-2800 Kongens Lyngby, Denmark}

\begin{abstract}
We address the long-standing unresolved problem concerning the $V$--parameter in a photonic crystal fiber (PCF). Formulate the parameter appropriate for a core-defect in a periodic structure we argue that the multi-mode cut-off occurs at a wavelength $\lambda^*$ which satisfies $V_{\rm PCF}(\lambda^*)=\pi$. Comparing to numerics and recent cut-off calculations we confirm this result.
\end{abstract}


\maketitle

In photonic crystal fibers (PCFs) an arrangement of air-holes running along the full length of the fiber provides the confinement and guidance of light. The air-holes of diameter $d$ are typically arranged in a triangular lattice\cite{knight1996} with a pitch $\Lambda$ (see insert in Fig.~\ref{fig2}), but {\it e.g.} honey-comb\cite{knight1998} and kagome\cite{benabid2002,kagome} arrangements are other options. By making a defect in the lattice, light can be confined and guided along the fiber axis. The guidance mechanism depends on the nature of the defect and the air-hole arrangement. For the triangular lattice with a silica-core light is confined by total-internal reflection\cite{knight1996} whereas for an air-core a photonic-bandgap confines light to the defect.\cite{cregan1999} For recent reviews we refer to Ref.~\onlinecite{russell2003} and references therein. 

Both type of PCFs have revealed surprising and novel optical properties. In this work we consider the silica-core PCF (see insert in Fig.~\ref{fig2}) which was the one first reported.\cite{knight1996} This structure provides the basis of a variety of phenomena including the endlessly single-mode behaviour,\cite{birks1997} large-mode area PCFs,\cite{knight1998el} as well as highly non-linear PCF with unique dispersion properties.\cite{mogilevtsev1998,ferrando2000,knight2000} 

Properties of standard fibers are often parametrized by the so-called $V$--parameter and the entire concept is very close to the heart of the majority of the optical fiber community (see {\it e.g.} Refs.~\onlinecite{snyder,ghatak1998}). The cut-off properties and the endlessly single-mode phenomena of PCFs can also be qualitatively understood within this framework.\cite{knight1996,birks1997,knight1998_josaa,knight1999,broeng1999} However, the proper choice of the correct length scale for the $V$--parameter has, until now, remained unsolved as well as the value of $V^*$ that marks the second-order cut-off. In this Letter we clarify this problem and also put recent work on multi-mode cut-off\cite{mortensen2002a,kuhlmey2002} into the context of the $V$--parameter.

\begin{figure}[b!]
\begin{center}
\epsfig{file=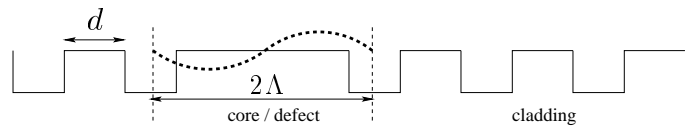, width=0.5\textwidth,clip}
\end{center}
\caption{Schematics of the cross-section of a PCF. The dashed line illustrates the field-amplitude of a second-order mode with a single node.}
\label{fig1}
\end{figure}

The tradition of parametrizing the optical properties in terms of the $V$--parameter stems from analysis of the step-index fiber (SIF). The SIF is characterized by the core radius $\rho$, the core index $n_c$, and the cladding index $n_{\rm cl}$ which all enter into the parameter $V_{\rm SIF}$ given by

\begin{equation}\label{VSIF}
V_{\rm SIF}(\lambda)= \frac{2\pi\rho}{\lambda}\sqrt{n_c^2-n_{\rm cl}^2}.
\end{equation} 
Because of its inverse dependence on the wavelength $\lambda$, this quantity is often referred to as the normalized frequency. However, in a more general context, this is somewhat misleading (especially if $n_c$ and/or $n_{\rm cl}$ has a strong wavelength dependence) and in this Letter we would like to emphasize a more physical interpretation. To do this, we first introduce the numerical aperture NA (or the angle of divergence $\theta$) given by

\begin{equation}
{\rm NA}=\sin\theta=\sqrt{n_c^2-n_{\rm cl}^2}
\end{equation}
which follows from use of Snell's law for critical incidence at the interface between the $n_c$ and $n_{\rm cl}$ regions (see {\it e.g.} Refs.~\onlinecite{snyder,ghatak1998}). Next, we introduce the free-space wave-number $k=2\pi/\lambda$ and its transverse projection $k_\perp = k \sin\theta$. The $V$--parameter can now simply be written as
\begin{equation}
V_{\rm SIF}= k_\perp\rho.
\end{equation} 
From this form it is obvious why the parameter carries information about the number of guided modes; the natural parameter describing the transverse intensity distribution is nothing but $ k_\perp\rho$. Furthermore, for the second-order cut-off wavelength $\lambda^*$ the usual value $V_{\rm SIF}(\lambda^*)=V_{\rm SIF}^*\simeq 2.405$ follows naturally from the solution of the first zero of the Bessel function, {\it i.e.} $J_0(V_{\rm SIF}^*)=0$.

In general, for wave-propagation in confined structures the number $ k_\perp\rho$ has a very central role. The transmission cross-section of a narrow slit\cite{montie1991} is an example and counterparts of the electro-magnetic problem can also be seen in {\it e.g.} electronic systems like the quantum-point contact where $ k_\perp\rho$ also determines the number of modes (see {\it e.g.} Ref.~\onlinecite{szafer1989}). In the context of PCFs it is also natural to consider a $V$--parameter which was done already in the seminal work by the Bath--group\cite{knight1996} and in the subsequent work on endlessly single-mode properties\cite{birks1997} and effective $V$--values.\cite{knight1998_josaa} However, in attempt of adopting Eq.~(\ref{VSIF}) to PCFs one is faced with the problem of choosing a value for $\rho$ and in Refs.~\onlinecite{birks1997,knight1998_josaa} it was emphasized that one may choose any transverse dimension. In this Letter, we point out that the problem is not a matter of defining a core-radius, but rather one should look for the natural length-scale of the problem; the air-hole pitch $\Lambda$. This choice was also suggested in Ref.~\onlinecite{birks1997} though considered an arbitrary choice. Regarding the second-order cut-off it was in Refs.~\onlinecite{knight1998_josaa} suggested that $V_{\rm PCF}^* \approx 2.5$ but it was also concluded that the arbitrary choice of the length scale means that the particular number for $V_{\rm PCF}^*$ also becomes somewhat arbitrary. In this Letter, we demonstrate that this is not the case and that a very simple and elegant solution exists.

To show this, we introduce the following $V$--parameter for a PCF
\begin{equation}\label{VPCF}
V_{\rm PCF}(\lambda)= \frac{2\pi\Lambda}{\lambda}\sqrt{n_c^2(\lambda)-n_{\rm cl}^2(\lambda)}
\end{equation}
where $n_c(\lambda)=c\beta/\omega$ is the ``core index'' associated with the effective index of the fundamental mode and similarly $n_{\rm cl}(\lambda)$ is the effective index of the fundamental space-filling mode in the triangular air-hole lattice. The second-order cut-off occurs at a wavelength $\lambda^*$ where the effective transverse wavelength $\lambda_\perp= 2\pi/k_\perp$ allows a mode with a single node (see schematics in Fig.~\ref{fig1}) to fit into the defect region, {\it i.e.} $\lambda_\perp^*\simeq 2\Lambda$. Writing Eq.~(\ref{VPCF}) in terms of $k_\perp$ the corresponding value of $V_{\rm PCF}^*$ easily follows

\begin{equation}\label{VPCF*}
V_{\rm PCF}^* =k_\perp^* \Lambda=\frac{2\pi}{ \lambda_\perp^*} \Lambda =\pi.
\end{equation}
Though this derivation may seem somewhat heuristic we shall compare to numerical results and show that the very central number $\pi$ is indeed the correct value.

\begin{figure}[t!]
\begin{center}
\epsfig{file=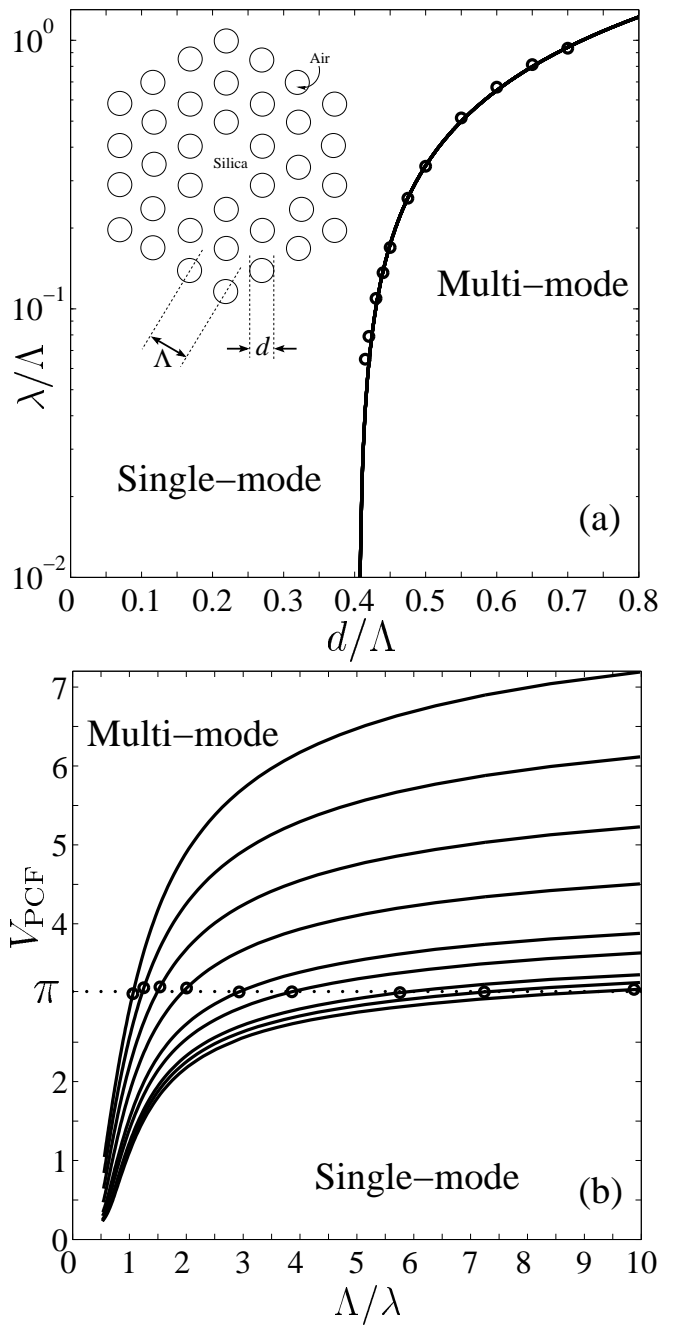, width=0.49\textwidth,clip}
\end{center}
\caption{Panel (a) shows the single/multi-mode phase diagram. The solid line shows the phase-boundary of Kuhlmey {\it et al.}\cite{kuhlmey2002} [Eq.~(\ref{cut-off})] and the circles indicate solutions to $V_{\rm PCF}(\lambda^*)=\pi$ [Eqs.~(\ref{VPCF},\ref{VPCF*})]. Panel (b) shows numerical results for PCFs with varying hole diameter ($d/\Lambda=0.43$, 0.44, 0.45, 0.475, 0.50, 0.55, 0.60, 0.65, and 0.70 from below). The full lines show results for the $V$--parameter [Eq.~(\ref{VPCF})], the circles indicate the corresponding cut-off wavelengths [Eq.~(\ref{cut-off})], and the dashed line shows $V_{\rm PCF}^*$ [Eq.~(\ref{VPCF*})].}
\label{fig2}
\end{figure}

For the numerical comparison we need to calculate both $V_{\rm PCF}(\lambda)$ and the second-order cut-off $\lambda^*$. For the $V$--parameter we use a fully-vectorial plane-wave method\cite{johnson2001} to calculate $n_c(\lambda)$ and $n_{\rm cl}(\lambda)$ for various air-hole diameters. For the material refractive index we use $n=1$ for the air-holes and $n=1.444$ for the silica. Ignoring the frequency dependence of the latter, the wave equation becomes scale-invariant~\cite{joannopoulos} and all the results to be presented can thus be scaled to the desired value of $\Lambda$. Regarding the cut-off, one of us recently suggested a phase diagram for the single and multi-mode operation regimes\cite{mortensen2002a} which was subsequently followed up in more detail by Kuhlmey {\it et al.}\cite{kuhlmey2002} From highly accurate multi-pole solutions of Maxwell's equations, it was numerically found that the single/multi-mode boundary can be accounted for by the expression\cite{kuhlmey2002}

\begin{equation}\label{cut-off}
\lambda^*/\Lambda\simeq \alpha (d/\Lambda-d^*/\Lambda)^{\gamma}.
\end{equation}
Here, $\alpha\simeq 2.80\pm 0.12 $, $\gamma\simeq 0.89\pm 0.02$, and $d^*/\Lambda \simeq 0.406$.
This phase-boundary is shown by the solid line in panel (a) of Fig.~\ref{fig2} and it has recently been confirmed experimentally based on cut-off measurements in various PCFs.\cite{folkenberg_preprint} For $d/\Lambda < d^*/\Lambda$ the PCF has the remarkable property of being so-called endlessly single-mode\cite{birks1997} and for $d/\Lambda> d^*/\Lambda$ the PCF supports a second-order mode at wavelengths $\lambda/\Lambda < \lambda^*/\Lambda$ and is single-mode for $\lambda/\Lambda > \lambda^*/\Lambda$.

In panel (b) of Fig.~\ref{fig2} we show numerical results for various values of $d/\Lambda$. The full lines show results for the $V$--parameter, Eq.~(\ref{VPCF}), the circles indicate the corresponding cut-off wavelengths, Eq.~(\ref{cut-off}), and the dashed line shows $V_{\rm PCF}^*$, Eq.~(\ref{VPCF*}). First of all we notice that the cut-off results of Kuhlmey {\it et al.},\cite{kuhlmey2002} Eq.~(\ref{cut-off}), agrees with a picture of a constant $V$--value $V_{\rm PCF}^*$ below which the PCF is single-mode. This similarity with SIFs indicate that the cut-off in SIFs and PCFs rely on the same basic physics. Furthermore, it is also seen that the cut-off points are in excellent agreement with the value $V_{\rm PCF}^*=\pi$, Eq.~(\ref{VPCF*}), and this also supports the idea of $\Lambda$ as the natural length scale for the $V$--parameter. We emphasize that the extremely small deviations from this value are within the accuracy of the coefficients in Eq.~(\ref{cut-off}). In panel (a) the data-points show cut-off results calculated from $V_{\rm PCF}(\lambda^*)=\pi$ and these results are in perfect agreement with the results of Kuhlmey {\it et al.}.\cite{kuhlmey2002}

In conclusion we have shown that the multi-mode cut-off in PCFs can be understood from a generalized $V$--parameter and that the single-mode regime is characterized by $V_{\rm PCF}< V_{\rm PCF}^*=\pi$.

\vspace{5mm}
N.~A. Mortensen is grateful to B.~T. Kuhlmey for stimulating discussions and M.~D. Nielsen and K.~P. Hansen acknowledge financial support by the Danish Academy of Technical Sciences. N.~A. Mortensen's e-mail address is nam@crystal-fibre.com.

\end{document}